# Experimental study and modeling of the lower-level controller of automated vehicle

Hua-Qing Liu, Shi-Teng Zheng, Rui Jiang, Junfang Tian, Ruidong Yan, Fang Zhang, Dezhao Zhang

*Abstract*—**Accurate modeling of lower-level controller plays an important role in the traffic flow of automated vehicles (AVs). However, there lacks enough attention with this respect. To address this issue, we conduct a field experiment with two vehicles that are equipped with developable autonomous driving system, where one can customize the upper-level control algorithm. Based on the field experimental data, a new lower-level control model is developed and compared with two widely used ones. The comparison results show that the proposed model outperforms the two previous models in capturing the observed actual acceleration, especially the troughs of the acceleration time series. Furthermore, theoretical analysis indicates that comparing with the proposed model, the two previous models significantly overestimate the stability region of the traffic flow of the AVs and the capacity of stable traffic flow. Our study is expected to further shed light on the importance of accurate lower-level control modeling.**

## I. Introduction

EMERGING automated vehicles (AVs) are expected to significantly improve the traffic performance in the near future. The modeling of longitudinal and lateral dynamics plays an important role in the control design of AVs [1]. Whereas longitudinal dynamics dominate the car-following process [2, 3], lateral dynamics mainly concern the lane-changing behavior of AVs [4, 5].

This paper focuses on the AV's longitudinal dynamics modeling. One approach is like modeling the conventional human-driven vehicles, which applies to the situation that the control algorithm of AV is unknown. Intelligent driver model (IDM) [6] is the most commonly used model. For example,

This work is supported in part by the Fundamental Research Funds for the Central Universities under Grant 2021JBZ107 and in part by the National Natural Science Foundation of China under Grant 72288101 and Grant 71931002.

H. Q. Liu, S. T. Zheng, R. Jiang and R. D. Yan are with the Key Laboratory of Transport Industry of Big Data Application Technologies for Comprehensive Transport, Ministry of Transport, Beijing Jiaotong University, Beijing 100044, China (e-mail: liuhuaqing@bjtu.edu.cn; stzheng@bjtu.edu.cn; jiangrui@bjtu.edu.cn; ruidongyan@yeah.net). *(Corresponding authors: Shi-Teng Zheng and Rui Jiang)*

J. F. Tian is with the Institute of Systems Engineering, College of Management and Economics, Tianjin University, Tianjin 300072, China (e-mail: jftian@tju.edu.cn).

F. Zhang and D. Z. Zhang are with the Beijing Idriverplus Technology Co., Ltd., Beijing 100044, China (e-mail: zhangfang@idriverplus.com; zhangdezhao@idriverplus.com)

Zhou et al. [7] developed a cooperative IDM and found that a higher AV penetration rate will smooth freeway oscillation and make the mixed traffic system much safer. An enhanced IDM is proposed in Ref.[8] to investigate the influence of two clustering strategies of Cooperative-AVs on the mixed traffic system. Li et al. [9] introduced accelerations of both leading AV and preceding one into IDM to investigate the rear-end collision risks. Shang and Stern [10] modeled the theorical adaptive cruise control (ACC) and commercial ACC vehicles based on the IDM with different parameter sets and studied the impacts on traffic stability and highway throughput.

Apart from IDM, other models are also used in AV modeling. Gunter et al. [11] tested three microscopic models based on experimentally collected data, and pointed out that the performances of IDM and the optimal velocity model with a relative velocity term (OVRV) are best and similar to each other, whereas the Gazis-Herman-Rothery model fails to capture all the ACC car-following dynamics. Furthermore, Gunter et al. [12] studied the string stability based on the calibrated OVRV, and found that all the tested commercial ACC systems are string unstable. Orosz et al. [13] focused on the robotic reaction times and showed that it will have short wave oscillations for any car-following model. The optimal velocity model is used to verify the conclusion that there are trade-offs between time delay and control gains. Mu et al. [14] adopted the desired safety margin model as an ACC velocity control method to investigate string stability and car-following safety. Yu et al. [15] conducted numerical simulations using the relative velocity difference model with memory in the ACC strategy to study the traffic stability and fuel consumptions.

Whereas most studies lack validation from empirical data because there is no AVs in the real traffic, some recent field experimental data were collected from commercial ACC vehicle (see e.g., [11, 12, 16-20]), which are expected to strength this modeling framework.

On the other hand, if the control algorithm of AV is known or one designs the control algorithm, a hierarchical modeling framework can be adopted [21]. The upper-level controller is designed to generate the command target, and the lower-level controller executes the command.

The command target usually has two types: (i) command acceleration [22-24] and (ii) command velocity [18, 25-27]. For example, Naus et al. [22] designed an ACC feedback controller, whereas Lidstrom et al. [23] and Ploeg et al. [24] designed a Cooperative-ACC (CACC) feedforward control to obtain the command acceleration. Milanés et al. [18] and Flores and Milanés [25] designed the command velocity based on the constant time gap car-following policy. In Refs.[28-30], the command target is generated via error tracking that aims to minimize the error between actual distance and desired distance. The learning-based method [31-36] and neural networks [37-42], as well as optimization method [43-45] have also been used in the design of upper-level control.

The lower-level controller determined by vehicle mechanical structure, also known as the vehicle model, achieves the command target by the throttle/brake maneuvers. Note that the command throttle/brake is often updated or compensated from command target via PID control, see, e.g., [46-48]. The most common lower-level controller is the first-order lag model [49-51]. The first-order lag model in frequency domain reads

$$G(s) = \frac{1}{T_d s + 1} \tag{1}$$

where $T_d$ denotes the time delay between the actual speed/acceleration and the command one.

Another lower-level control model is the second-order response model with time delay,

$$G(s) = \frac{m_1 s + K_0}{m_2 s^2 + m_3 s + 1} e^{-T_d s} \tag{2}$$

where $K_0$ denotes the static gain; $m_1$ is a nonnegative parameter, $m_2$ and $m_3$ are the positive parameters in the controller, respectively. In Refs.[18, 25], some field experiments are carried out and the second-order response model without zero point (i.e., $m_1 = 0$) is calibrated. The results showed that the second-order response model exhibits high accuracy in reproducing the experimental results.

However, in the field experiments [18, 25], only step speed signal has been used as the command target in the acceleration and braking maneuvers. In real traffic situation, the vehicle speed often experiences oscillations, and these oscillations could possibly cause an accumulation of the mechanical transmission errors.

Motivated by the fact, recently, we conduct a field experiment using two developable AVs, aiming to establish a more accurate lower-level control model under speed oscillation situation.

The rest of this paper is organized as follows. Section 2 reports the experiment setup. Section 3 models the lower-level controller of AV, presents the calibration and validation results of the proposed model and also compares it with the aforementioned models. Section 4 conducts the stability analysis for these models. Finally, conclusions are given in Section 5.

## II. EXPERIMENT SETUP

The experiment was performed on March 29, 2021. Two vehicles equipped with developable autonomous driving systems were used in the experiment, see Fig. 1. We conduct car-following test, in which the leading vehicle tracks a pre-set speed profile, and the following vehicle follows the leading one under the given upper-level longitudinal control algorithm. The vehicles moved straight ahead and did not change lane. The steering wheel of the two vehicles is controlled by human driver. Both vehicles are front-wheel drive ones reconfigured from CHANGAN Auto. The equipped control computer calculates the speed, acceleration and yaw rate according to the target speed or tracking speed from lidar, GPS and other on-board vehicle sensors. Initially the two vehicles are stopped with a standstill spacing $G_{min}$ (rear axle of ego vehicle to the center of preceding vehicle). The data are collected with the frequency of 20 Hz.

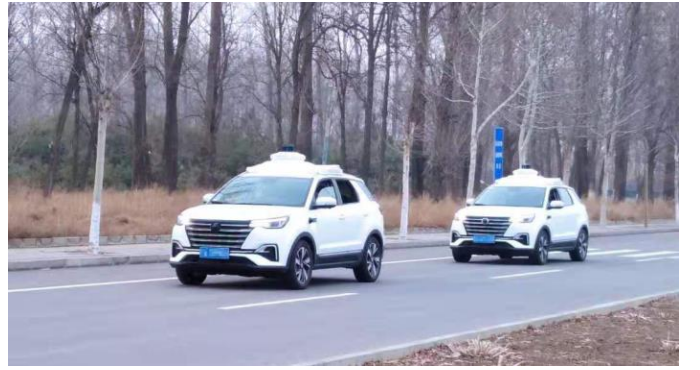

Fig. 1. The experimental vehicles.

In our study, the constant time gap car-following policy is used in the upper-level longitudinal control to generate the command acceleration target, which reads

$$\begin{aligned} a_{cmd}(t) = {} & k_g \left( x_{n-1}(t) - x_n(t) - T_g v_n(t) - G_{min} \right) \\ & + k_v \left( v_{n-1}(t) - v_n(t) \right) \end{aligned} \tag{3}$$

where $a_{cmd}(t)$ is the command acceleration; $k_v$ and $k_g$ are the control gains; $T_g$ is the constant time gap. $v_n(t)$ and $x_n(t)$ are the velocity and location of the $n$th vehicle, respectively. $n-1$ denotes the subscript of preceding vehicle. In the experiment, 21 runs were carried out, in which the parameters of $k_g$ and $T_g$ in each run are listed in Table 1. $k_v$ is always set to 0. The standstill spacing $G_{min}$ is set to 9.5 m. We set such a large value of $G_{min}$ to ensure the safety of vehicles.

TABLE I
DETAILS OF THE EXPERIMENT RUNS IN CHRONOLOGICAL ORDER.

| Run number | $k_g$ (s$^{-2}$) | $T_g$ (s) | Run number | $k_g$ (s$^{-2}$) | $T_g$ (s) |
|---|---|---|---|---|---|
| 1 | 0.7 | 2.0 | 12 | 0.6 | 2.0 |
| 2 | 0.7 | 2.0 | 13 | 0.6 | 2.0 |
| 3 | 0.5 | 1.8 | 14 | 0.6 | 2.0 |
| 4 | 0.5 | 1.8 | 15 | 0.5 | 2.0 |
| 5 | 0.5 | 1.8 | 16 | 0.5 | 2.0 |
| 6 | 0.5 | 1.8 | 17 | 0.5 | 2.0 |
| 7 | 0.5 | 1.8 | 18 | 0.5 | 2.0 |
| 8 | 0.6 | 2.0 | 19 | 0.4 | 2.0 |
| 9 | 0.6 | 2.0 | 20 | 0.4 | 2.0 |
| 10 | 0.6 | 2.0 | 21 | 0.4 | 2.0 |
| 11 | 0.6 | 2.0 | — | — | — |

## III. LOWER-LEVEL CONTROLLER OF AUTOMATED VEHICLE

To capture the vehicle dynamics of AVs precisely, the mechanical control in lower level should be accurately modeled. The lower-level controller transmits the command acceleration from upper-level control to the actual acceleration,

which reads

$$A(s) = G(s) A_{cmd}(s) \qquad (4)$$

where $A_{cmd}(s)$ and $A(s)$ denote the command acceleration and actual one in frequency domain, respectively; $G(s)$ denotes the transfer function from $A_{cmd}(s)$ to $A(s)$.

This section develops an improved lower-level control model, and compare it with the two aforementioned ones, i.e., Eq.(1) and Eq.(2).

### A. Improved lower-level controller

To model the motion of AV, the second-order response controller, i.e., Eq.(2), is implemented as a basis in the lower-level controller. Different from Refs.[18, 25], $m_1$ is set as a positive parameter in the controller, and $-K_0/m_1$ thus means one zero point in the model.

To track command acceleration better and correct the output, an inner feedback controller is introduced to improve the lower-level controller $G(s)$. Thus, the controller output will make a real-time correction over the actual acceleration in the inner feedback loop. This turns from tracking command acceleration to tracking reference acceleration $A_n^{ref}$. After introducing the inner feedback controller, the lower-level controller changes from $G(s)$ to $G_{fb}(s)$ as follows:

$$G_{fb}(s) = \frac{G(s)}{1 - K \cdot G(s)} \qquad (5)$$

Fig. 2 shows the AV control structure with $G_{fb}(s)$. The status of preceding vehicle $A_{n-1}(s)$ can be captured by the on-board vehicle sensors and finally is converted to the status of ego vehicle $A_n(s)$ through AV system.

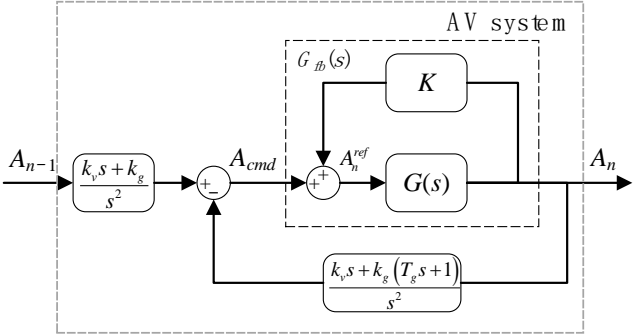

Fig. 2. AV control structure.

### B. Model calibration

The genetic algorithm (GA) is employed to solve the calibration problem in the Optimization Toolbox of MATLAB. The fitness function is set to minimize the mean square error (MSE) between the simulated actual accelerations and experimental ones, which reads

$$\min \frac{1}{M} \sum_j \frac{1}{T} \sum_t \left( a_{t,j}^{sim} - a_{t,j}^{exp} \right)^2 \qquad (6)$$

where $a_{t,j}^{sim}$ and $a_{t,j}^{exp}$ denote the actual accelerations in the simulation and experiment at time $t$ in Run $j$, respectively; $T$

denotes the time length of the Run, and $M$ denotes the number of runs.

In the simulations, $\Delta t = 0.05$ s. We randomly choose 10 runs used for calibration, and the rest of runs will be used for model validation. The calibrated model parameters of lower-level controller are shown in Table 2.

TABLE II
CALIBRATED MODEL PARAMETERS OF LOWER-LEVEL CONTROLLER.

| Parameter | Unit | Lower bound | Upper bound | Calibrated | | |
| --- | --- | --- | --- | --- | --- | --- |
| | | | | First-order lag model | Second-order response model | The proposed model |
| $m_1$ | s | 0 | Inf | — | — | 6.7893 |
| $m_2$ | s² | 0 | Inf | — | 0.0445 | 1.2824 |
| $m_3$ | s | 0 | Inf | — | 0.1305 | 8.8157 |
| $K_0$ | 1 | 0 | Inf | — | 0.7292 | 0.3479 |
| $T_d$ | s | 0 | 3 | 1.0758 | 0.7796 | 0.7903 |
| $K$ | 1 | –Inf | Inf | — | — | 0.1008 |

'—' denotes not applicable.

### C. Model validation and comparison

In this subsection, the model calibration and validation results are presented. One can see in Figs. 3 and 4 that simulation results of the proposed model are in good agreement with the field data, especially the troughs of acceleration. It is worth to mention that although the first-order lag model is more popular since its simple structure is convenient for theoretical analysis, it cannot accurately capture the peaks and troughs of speed oscillation. The calibration MSE errors of actual acceleration are reported in Table 3. One can see that the average calibration errors are 0.0398, 0.0551 and 0.1232 (m/s²)² for the proposed model, the second-order response model and the first-order lag model, respectively. The validation MSE errors of actual acceleration are reported in Table 4. The average validation errors are 0.0871, 0.1213 and 0.2587 (m/s²)², respectively. Therefore, the proposed model outperforms the first-order lag model and second-order response model.

We also would like to mention that if we set $m_1 = 0$ as in Refs.[18, 25], then calibration and validation errors become almost the same as the second-order response model, which are 0.0544 and 0.1191 (m/s²)², respectively.

TABLE III
THE CALIBRATION MSE ERRORS OF ACTUAL ACCELERATION FOR THE FIRST-ORDER LAG MODEL, THE SECOND-ORDER RESPONSE MODEL AND THE PROPOSED MODEL (UNIT: (M/S²)²).

| Run No. | 3 | 4 | 5 | 9 | 10 |
| --- | --- | --- | --- | --- | --- |
| The proposed model | 0.03 | 0.03 | 0.05 | 0.04 | 0.05 |
| Second-order response model | 0.04 | 0.04 | 0.07 | 0.06 | 0.07 |
| First-order lag model | 0.09 | 0.08 | 0.14 | 0.17 | 0.17 |
| Run No. | 11 | 15 | 16 | 17 | 19 |
| The proposed model | 0.06 | 0.04 | 0.03 | 0.03 | 0.05 |
| Second-order response model | 0.07 | 0.05 | 0.04 | 0.04 | 0.06 |
| First-order lag model | 0.18 | 0.12 | 0.09 | 0.09 | 0.12 |

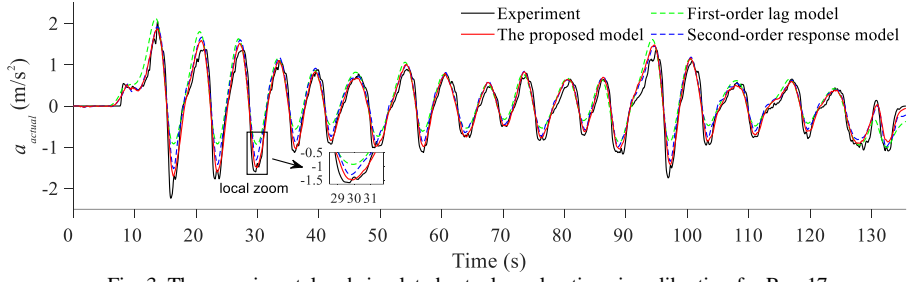

Fig. 3. The experimental and simulated actual accelerations in calibration for Run 17.

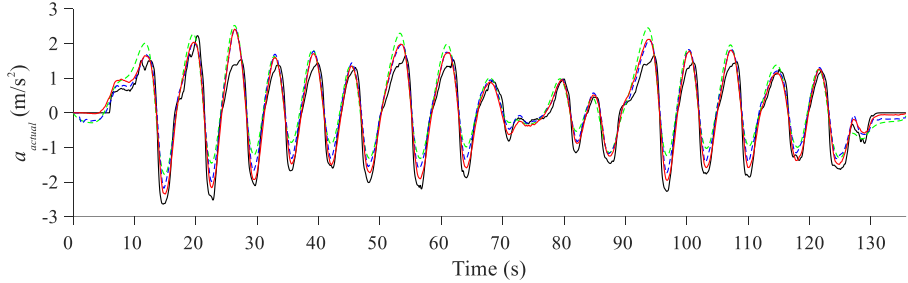

Fig. 4. The experimental and simulated actual accelerations in validation for Run 6.

TABLE IV
THE VALIDATION MSE ERRORS OF ACTUAL ACCELERATION FOR THE
FIRST-ORDER LAG MODEL, THE SECOND-ORDER RESPONSE MODEL AND THE
PROPOSED MODEL (UNIT: (M/S²)²).

| Run No. | 1 | 2 | 6 | 7 | 8 | 12 |
|---|---|---|---|---|---|---|
| The proposed model | 0.11 | 0.10 | 0.08 | 0.16 | 0.07 | 0.06 |
| Second-order response model | 0.16 | 0.13 | 0.13 | 0.22 | 0.09 | 0.08 |
| First-order lag model | 0.37 | 0.30 | 0.25 | 0.41 | 0.23 | 0.19 |
| Run No. | 13 | 14 | 18 | 20 | 21 | — |
| The proposed model | 0.07 | 0.07 | 0.06 | 0.07 | 0.11 | — |
| Second-order response model | 0.09 | 0.10 | 0.09 | 0.10 | 0.15 | — |
| First-order lag model | 0.22 | 0.24 | 0.18 | 0.19 | 0.27 | — |

## IV. STABILITY ANALYSIS

Although the improvement of the proposed model seems not striking in terms of MSE error, there is significant impact on the traffic flow stability and capacity. To clarify this issue, we conduct the stability analysis in this section.

Local stability and string stability [52] are the basic but important characteristic of a traffic system. Local stability means that the disturbance of the follower will decrease with time. String stability is platoon stability or asymptotic stability, which means that the propagation of disturbances attenuates in a string of vehicles.

In Fig. 2, the AV system is designed as follows.

$$A_n(s) = H(s) A_{n-1}(s) \qquad (7)$$

The transfer function $H(s)$ that integrates the upper-level controller Eq.(3) and the lower-level controller Eq.(5) in frequency domain is given by

$$H(s) = \frac{(k_v s + k_g) G_{fb}(s)}{s^2 + G_{fb}(s)\big((k_v + T_g k_g)s + k_g\big)} \qquad (8)$$

Firstly, the second-order Padé approximation is adopted to estimate the time delay $T_d$. $G(s)$ can be given by

$$G(s) = \frac{G_a}{G_b} \qquad (9)$$

where

$$\begin{cases} G_a = (m_1 s + K_0)\left[1 - p_1 T_d s + p_2 (T_d s)^2\right] \\ G_b = (m_2 s^2 + m_3 s + 1)\left[1 + p_1 T_d s + p_2 (T_d s)^2\right] \end{cases} \qquad (10)$$

and $p_1 = \frac{1}{2}$, $p_2 = \frac{1}{12}$ .

### A. Local Stability

The characteristic equation of $H(s)$ is given by

$$s^2 + G_{fb}(s)\big((k_v + T_g k_g)s + k_g\big) = 0 \qquad (11)$$

Substituting Eqs.(5) and (9) into Eq.(11), one has

$$(G_b - KG_a)s^2 + G_a(\mu s + k_g) = 0 \qquad (12)$$

where $\mu = k_v + T_g \cdot k_g$.

Substituting Eq.(10) into Eq.(12), one has

$$a_1 s^6 + a_2 s^5 + a_3 s^4 + a_4 s^3 + a_5 s^2 + a_6 s + a_7 = 0 \qquad (13)$$

where

$$\begin{cases} a_1 = m_2 T_d^2 p_2 \\ a_2 = -T_d^2 K m_1 p_2 + T_d^2 m_3 p_2 + T_d m_2 p_1 \\ a_3 = -K K_0 T_d^2 p_2 + K m_1 T_d p_1 + T_d^2 p_2 \\ \qquad + m_3 T_d p_1 + m_2 + m_1 T_d^2 p_2 \mu \\ a_4 = K K_0 T_d p_1 + T_d p_1 - K m_1 + m_3 \\ \qquad + \left(K_0 T_d^2 p_2 - T_d m_1 p_1\right)\mu + m_1 T_d^2 p_2 k_g \\ a_5 = -K K_0 + \left(-K_0 T_d p_1 + m_1\right)\mu \\ \qquad + \left(K_0 T_d^2 p_2 - T_d m_1 p_1\right)k_g + 1 \\ a_6 = K_0 \mu + \left(-K_0 T_d p_1 + m_1\right)k_g \\ a_7 = K_0 k_g \end{cases} \qquad (14)$$

To ensure local stability, all roots of characteristic equation should have no positive real part. Based on Routh-Hurwitz stability criterion, Routh Matrix constructed by coefficients of characteristic equation can be utilized to determine local stability, see Table 5.

TABLE V
THE ROUTH MATRIX.

| $s^6$ | $a_1$ | $a_3$ | $a_5$ | $a_7$ |
|---|---|---|---|---|
| $s^5$ | $a_2$ | $a_4$ | $a_6$ | 0 |
| $s^4$ | $b_1 = \frac{a_2 a_3 - a_1 a_4}{a_2}$ | $b_2 = \frac{a_2 a_5 - a_1 a_6}{a_2}$ | $b_3 = \frac{a_2 a_7 - a_1 \cdot 0}{a_2}$ | 0 |
| $s^3$ | $c_1 = \frac{b_1 a_4 - a_2 b_2}{b_1}$ | $c_2 = \frac{b_1 a_6 - a_2 b_3}{b_1}$ | 0 | |
| $s^2$ | $d_1 = \frac{c_1 b_2 - b_1 c_2}{c_1}$ | $d_2 = \frac{c_1 b_3 - b_1 \cdot 0}{c_1}$ | | |
| $s^1$ | $e_1 = \frac{d_1 c_2 - c_1 d_2}{d_1}$ | 0 | | |
| $s^0$ | $f_1 = \frac{e_1 d_2 - d_1 \cdot 0}{e_1}$ | | | |

Hence, the criterion for local stability of the proposed model is obtained, which requires that

(i) the coefficient of characteristic equation should be larger than 0:

$$a_i > 0 \tag{15a}$$

(ii) the elements in the first column of Routh Matrix should be larger than 0:

$$X_{i,1} > 0 \tag{15b}$$

where $X_{i,j}$ denotes the element in the Routh Matrix; $i$ and $j$ denote the row index and column index of matrix, respectively.

### B. String Stability

To ensure the string stability that small disturbance attenuates along the vehicle platoon, the magnitude of $|H(j\omega)|$ should satisfy $|H(j\omega)| = \sqrt{H(j\omega)H(-j\omega)} < 1$ for all positive $\omega$.

The transfer function Eq.(8) can be reformulated as

$$H(s) = \frac{b_1' s^4 + b_2' s^3 + b_3' s^2 + b_4' s + b_5'}{a_1 s^6 + a_2 s^5 + a_3 s^4 + a_4 s^3 + a_5 s^2 + a_6 s + a_7} \tag{16}$$

where

$$\begin{cases} b_1' = T_d^2 k_v m_1 p_2 \\ b_2' = K_0 T_d^2 k_v p_2 + T_d^2 k_g m_1 p_2 - T_d k_v m_1 p_1 \\ b_3' = K_0 T_d^2 k_g p_2 - K_0 T_d k_v p_1 - T_d k_g m_1 p_1 + k_v m_1 \\ b_4' = -K_0 T_d k_g p_1 + K_0 k_v + k_g m_1 \\ b_5' = K_0 k_g \end{cases} \tag{17}$$

Hence, string stability condition can be calculated as

$$c_1' \omega^{12} + c_2' \omega^{10} + c_3' \omega^8 + c_4' \omega^6 + c_5' \omega^4 + c_6' \omega^2 + c_7' > 0 \tag{18}$$

where

$$\begin{cases} c_1' = a_1^2 \\ c_2' = a_2^2 - 2a_1 a_3 \\ c_3' = 2a_1 a_5 - 2a_2 a_4 + a_3^2 - b_1'^2 \\ c_4' = -2a_1 a_7 + 2a_2 a_6 - 2a_3 a_5 + 2b_1' b_3' + a_4^2 - b_2'^2 \\ c_5' = 2a_3 a_7 - 2a_4 a_6 - 2b_1' b_5' + 2b_2' b_4' + a_5^2 - b_3'^2 \\ c_6' = -2a_5 a_7 + 2b_3' b_5' + a_6^2 - b_4'^2 \\ c_7' = a_7^2 - b_5'^2 \end{cases} \tag{19}$$

Since $a_7 = b_5'$, one has $c_7' = 0$. Therefore, Eq.(18) can be simplified as

$$c_1' \omega^{10} + c_2' \omega^8 + c_3' \omega^6 + c_4' \omega^4 + c_5' \omega^2 + c_6' > 0 \tag{20}$$

### C. The stability region

Fig. 5 compares the stability region of the proposed model with that of the second-order response model. One can see that the two models exhibit the similar features. At a given $k_v$, there exists the thresholds for both $k_g$ and $T_g$. Beyond the thresholds of $k_g$ or below the thresholds of $T_g$, traffic flow is always unstable. With the increase of $k_v$, the stable region shrinks. In particular, beyond the threshold of $k_v = 0.8085$ s$^{-1}$, traffic flow is unstable for the whole concerned parameter range (i.e., $T_g \in [0, 15]$ s) in the second-order response model. The threshold of $k_v$ in the proposed model is a little smaller, which is 0.7395 s$^{-1}$. Moreover, with the increase of $k_v$, the threshold of $k_g$ decreases, whereas the threshold of $T_g$ first decreases and then increases sharply, see Fig. 6. Therefore, comparing with the proposed model, the second-order response model significantly overestimates the traffic flow stability.

Moreover, the time gap $T_g$ is closely related to the traffic flow capacity. As a result, the second-order response model significantly overestimates the capacity of stable traffic flow. The minimum $T_g$ of stable traffic flow is 1.9 s in the second-order response model, which corresponds to capacity 1687.5 vehicle/h. However, in the propose model, the minimum $T_g$ is 3.5 s, which corresponds to the capacity 964.3 vehicle/h and is 42.86% lower than that in the second-order response model.

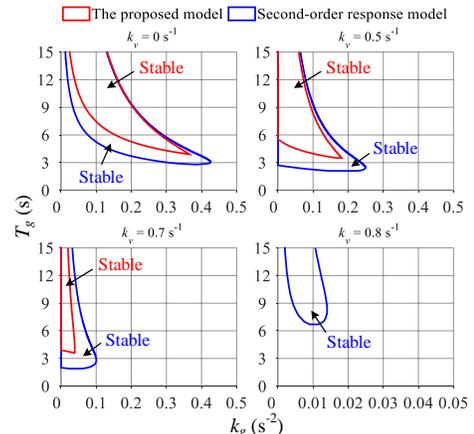

Fig. 5. The comparison of the stability region between the proposed model and the second-order response model.

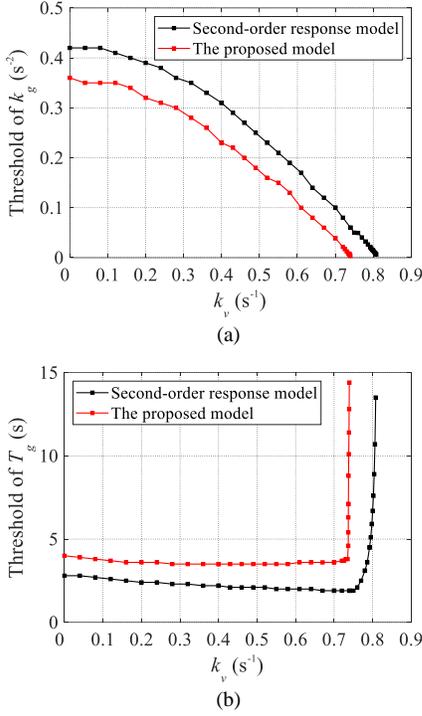

(a)

(b)

Fig. 6. The comparisons of threshold of (a) $k_g$ and (b) $T_g$ between the proposed model and the second-order response model.

Fig. 7 compares the stability region of the proposed model with that of the first-order lag model[1]. One can see that the traffic stability property of the first-order lag model exhibits striking difference from the proposed model. At a given $k_v$, for the first-order lag model there does not exist a threshold of $k_g$ but still exists a threshold of $T_g$. Fig. 8 shows the dependence of the threshold of $T_g$ on $k_v$. Comparing with the proposed model, the first-order lag model also significantly overestimates the traffic flow stability and the capacity of stable traffic flow. The minimum $T_g$ of stable traffic flow is 2.2 s in the first-order lag model, which corresponds to capacity 1479.5 vehicle/h and is 53.42% larger than that of the proposed model.

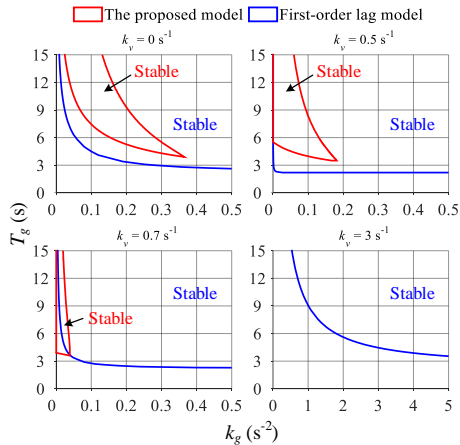

Fig. 7. The comparison of the stability region between the proposed model and the first-order lag model.

[1] The local and string stability conditions of the first-order lag model are reported in Appendix.

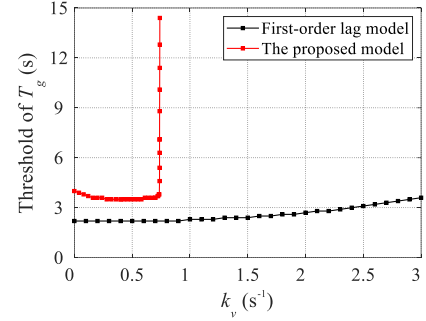

Fig. 8. The comparison of threshold of $T_g$ between the proposed model and the first-order lag model.

## V. Conclusion

In recent years, the field experimental measurements on automated vehicles are becoming increasingly available from around the world, which provides a new perspective to study the traffic flow features arising from this emerging technology. On the other hand, modeling and simulation are also useful tools to capture the AV longitudinal and lateral dynamics and explain the traffic flow characteristic, in which the accuracy of AV control model naturally plays an important role.

However, for the hierarchical modeling framework, most studies concentrate on the design of upper-level controller, and the lower-level control modeling lacks enough attention. The inaccuracy of the lower-level control model will lead to large deviation between theoretical analysis and field runs. To address this issue, we conducted a car-following experiment with two developable AVs, where the upper-level control logic can be customized and the lower-level control can thus be studied separately.

In the experiment, the command acceleration from upper-level controller and the actual acceleration from lower-level controller are collected for model identification. To accurately predict the actual acceleration, a new lower-level control model is developed by introducing feedback control into the second-order response model. The proposed model outperforms the first-order lag model and the second-order response model. The average calibration errors are 0.0398, 0.0551 and 0.1232 $(m/s^2)^2$ for the proposed model, the first-order lag model and second-order response model, respectively. The average validation errors are 0.0871, 0.1213 and 0.2587 $(m/s^2)^2$, respectively.

Furthermore, we perform stability analysis of the models. It is shown that comparing with the proposed model, the first-order lag model and the second-order response model significantly overestimate the stability region of the traffic flow of the AVs and thus the capacity of stable traffic flow.

In the future work, more field experiments with various oscillation scenarios and different vehicles are expected to be carried out to examine the proposed model. Moreover, although the proposed model outperforms the previous ones, there is still nontrivial deviation between simulation results and experimental ones. One possible reason is that acceleration and braking correspond to different lower-level control model. Further investigations are needed in the future work.

APPENDIX. THE STABILITY CONDITION OF THE FIRST-ORDER LAG MODEL.

For the AV model that integrates the upper-level controller Eq.(3) and the lower-level controller Eq.(1), the transfer function $H(s)$ in frequency domain can be given by

$$H(s) = \frac{k_v s + k_g}{T_d s^3 + s^2 + \left(k_v + k_g T_g\right)s + k_g} \qquad (A.1)$$

Hence the characteristic equation is

$$T_d s^3 + s^2 + \left(k_v + k_g T_g\right)s + k_g = 0 \qquad (A.2)$$

Based on Routh-Hurwitz stability criterion, the criterion for local stability can be given by

$$k_v + k_g \left(T_g - T_d\right) > 0 \qquad (A.3)$$

Considering $|H(s)| < 1$, the criterion for string stability can be given by

$$\begin{cases} k_v + T_g k_g \leq \dfrac{1}{2T_d} \\ T_g k_v + \dfrac{T_g^2 k_g}{2} > 1 \end{cases} \text{ or } \begin{cases} k_v + T_g k_g \geq \dfrac{1}{2T_d} \\ \left(k_v - \dfrac{1}{2T_d}\right)^2 < \left(\dfrac{T_g}{T_d} - 2\right)k_g \end{cases} \qquad (A.4)$$

## REFERENCES

[1] P. Liu, U. Ozguner, and Y. Zhang, "Distributed MPC for cooperative highway driving and energy-economy validation via microscopic simulations," *Transp. Res. C: Emerg. Technol.*, vol. 77, pp. 80-95, Apr 2017.

[2] R. Kianfar *et al.*, "Design and experimental validation of a cooperative driving system in the grand cooperative driving challenge," *IEEE T Intell. Transp. Sys.*, vol. 13, no. 3, pp. 994-1007, Sep 2012.

[3] J. Ploeg, D. P. Shukla, N. van de Wouw, and H. Nijmeijer, "Controller synthesis for string stability of vehicle platoons," *IEEE T Intell. Transp. Sys.*, vol. 15, no. 2, pp. 854-865, Apr 2014.

[4] Y. Luo, Y. Xiang, K. Cao, and K. Li, "A dynamic automated lane change maneuver based on vehicle-to-vehicle communication," *Transp. Res. C: Emerg. Technol.*, vol. 62, pp. 87-102, Jan 2016.

[5] B. Morris, A. Doshi, M. Trivedi, and Ieee, "Lane change intent prediction for driver assistance: On-road design and evaluation," in *2011 IEEE Intelligent Vehicles Symposium*, 2011, pp. 895-901.

[6] M. Treiber, A. Hennecke, and D. Helbing, "Congested traffic states in empirical observations and microscopic simulations," *Phys. Rev. E*, vol. 62, no. 2, pp. 1805-1824, Aug 2000.

[7] M. Zhou, X. Qu, and S. Jin, "On the impact of cooperative autonomous vehicles in improving freeway merging: A modified intelligent driver model-based approach," *IEEE T Intell. Transp. Sys.*, vol. 18, no. 6, pp. 1422-1428, Jun 2017.

[8] Z. Zhong, E. E. Lee, M. Nejad, and J. Lee, "Influence of CAV clustering strategies on mixed traffic flow characteristics: An analysis of vehicle trajectory data," *Transp. Res. C: Emerg. Technol.*, vol. 115, Jun 2020, Art no. 102611.

[9] Y. Li, H. Wang, W. Wang, L. Xing, S. Liu, and X. Wei, "Evaluation of the impacts of cooperative adaptive cruise control on reducing rear-end collision risks on freeways," *Accid. Anal. Prev.*, vol. 98, pp. 87-95, Jan 2017.

[10] M. Shang and R. E. Stern, "Impacts of commercially available adaptive cruise control vehicles on highway stability and throughput," *Transp. Res. C: Emerg. Technol.*, vol. 122, Jan 2021, Art no. 102897.

[11] G. Gunter, R. Stern, D. B. Work, and Ieee, "Modeling adaptive cruise control vehicles from experimental data: Model comparison," in *IEEE Intelligent Transportation Systems Conference (IEEE-ITSC)*, Auckland, New Zealand, 2019, pp. 3049-3054.

[12] G. Gunter, D. Gloudemans, R. E. Stern, S. Mcquade, and D. B. Work, "Are commercially implemented adaptive cruise control systems string stable?," *IEEE T Intell. Transp. Sys.*, vol. PP, no. 99, pp. 1-12, 2020.

[13] G. Orosz, J. Moehlis, and F. Bullo, "Robotic reactions: Delay-induced patterns in autonomous vehicle systems," *Phys. Rev. E*, vol. 81, no. 2, Feb 2010, Art no. 025204.

[14] M. Mu, J. Zhang, C. Wang, J. Zhang, and C. Yang, "String stability and platoon safety analysis of a new car-following model considering a stabilization strategy," *IEEE Access*, vol. 9, pp. 111336-111345, 2021.

[15] S. Yu, J. Tang, and Q. Xin, "Relative velocity difference model for the car-following theory," *Nonlinear Dyn.*, vol. 91, no. 3, pp. 1415-1428, Feb 2018.

[16] M. Makridis, K. Mattas, A. Anesiadou, and B. Ciuffo, "OpenACC. An open database of car-following experiments to study the properties of commercial acc systems," *Transp. Res. C: Emerg. Technol.*, vol. 125, Apr 2021, Art no. 103047.

[17] G. Gunter, C. Janssen, W. Barbour, R. E. Stern, and D. B. Work, "Model-based string stability of adaptive cruise control systems using field data," *IEEE T Intell. Veh.*, vol. 5, no. 1, pp. 90-99, 2020.

[18] V. Milanés, S. E. Shladover, J. Spring, C. Nowakowski, H. Kawazoe, and M. Nakamura, "Cooperative adaptive cruise control in real traffic situations," *IEEE T Intell. Transp. Sys.*, vol. 15, no. 1, pp. 296-305, Feb 2014.

[19] T. Li, D. Chen, H. Zhou, J. Laval, and Y. Xie, "Car-following behavior characteristics of adaptive cruise control vehicles based on empirical experiments," *Transp. Res. B: Method.*, vol. 147, pp. 67-91, May 2021.

[20] X. Shi and X. Li, "Constructing a fundamental diagram for traffic flow with automated vehicles: Methodology and demonstration," *Transp. Res. B: Method.*, vol. 150, pp. 279-292, Aug 2021.

[21] R. Rajamani and S. E. Shladover, "An experimental comparative study of autonomous and co-operative vehicle-follower control systems," *Transp. Res. C: Emerg. Technol.*, vol. 9, no. 1, pp. 15-31, Feb 2001.

[22] G. J. L. Naus, R. P. A. Vugts, J. Ploeg, M. J. G. van de Molengraft, and M. Steinbuch, "String-stable CACC design and experimental validation: A frequency-domain approach," *IEEE T Veh. Technol.*, vol. 59, no. 9, pp. 4268-4279, Nov 2010.

[23] K. Lidstrom *et al.*, "A modular CACC system integration and design," *IEEE T Intell. Transp. Sys.*, vol. 13, no. 3, pp. 1050-1061, Sep 2012.

[24] J. Ploeg, N. van de Wouw, and H. Nijmeijer, "L-p string stability of cascaded systems: Application to vehicle platooning," *IEEE T Contr. Sys. Technol.*, vol. 22, no. 2, pp. 786-793, Feb 2014.

[25] C. Flores and V. Milanés, "Fractional-order-based ACC/CACC algorithm for improving string stability," *Transp. Res. C: Emerg. Technol.*, vol. 95, pp. 381-393, Oct 2018.

[26] H. Hao and P. Barooah, "On achieving size-independent stability margin of vehicular lattice formations with distributed control," *IEEE T Automat. Contr.*, vol. 57, no. 10, pp. 2688-2694, Oct 2012.

[27] R. Horowitz and P. Varaiya, "Control design of an automated highway system," *Proc. IEEE*, vol. 88, no. 7, pp. 913-925, Jul 2000.

[28] S. Oencue, J. Ploeg, N. van de Wouw, and H. Nijmeijer, "Cooperative adaptive cruise control: Network-aware analysis of string stability," *IEEE T Intell. Transp. Sys.*, vol. 15, no. 4, pp. 1527-1537, Aug 2014.

[29] E. Kayacan, "Multi-objective $H_\infty$ control for string stability of cooperative adaptive cruise control systems," *IEEE T Intell. Veh.*, vol. 2, no. 1, pp. 52-61, 2017.

[30] Y. Abou Harfouch, S. Yuan, and S. Baldi, "An adaptive switched control approach to heterogeneous platooning with intervehicle communication losses," *IEEE T Control. Netw. Syst.*, vol. 5, no. 3, pp. 1434-1444, Sep 2018.

[31] I. Rasheed, F. Hu, and L. Zhang, "Deep reinforcement learning approach for autonomous vehicle systems for maintaining security and safety using LSTM-GAN," *Veh. Commun.*, vol. 26, Dec 2020, Art no. 100266.

[32] C. You, J. Lu, D. Filev, and P. Tsiotras, "Advanced planning for autonomous vehicles using reinforcement learning and deep inverse reinforcement learning," *Robot. Auton. Syst.*, vol. 114, pp. 1-18, Apr 2019.

[33] R. P. Bhattacharyya, D. J. Phillips, B. Wulfe, J. Morton, A. Kuefler, and M. J. Kochenderfer, "Multi-agent imitation learning for driving simulation," in *25th IEEE/RSJ International Conference on Intelligent Robots and Systems (IROS)*, Madrid, Spain, 2018, pp. 1534-1539.

[34] X. Di and R. Shi, "A survey on autonomous vehicle control in the era of mixed-autonomy: From physics-based to AI-guided driving policy learning," *Transp. Res. C: Emerg. Technol.*, vol. 125, Apr 2021, Art no. 103008.

[35] X. Wang, R. Jiang, L. Li, Y. Lin, X. Zheng, and F.Y. Wang, "Capturing car-following behaviors by deep learning," *IEEE T Intell. Transp. Sys.*, vol. 19, no. 3, pp. 910-920, Mar 2018.

[36] Z. Qi, T. Wang, J. Chen, D. Narang, Y. Wang, and H. Yang, "Learning-based path planning and predictive control for autonomous vehicles with low-cost positioning," *IEEE T Intell. Veh.*, Jan 2022. DOI: 10.1109/TIV.2022.3146972.

[37] D. Zhao, Z. Hu, Z. Xia, C. Alippi, Y. Zhu, and D. Wang, "Full-range adaptive cruise control based on supervised adaptive dynamic programming," *Neurocomputing*, vol. 125, pp. 57-67, Feb 11 2014.

[38] F. Hegedus, T. Becsi, S. Aradi, and G. Galdi, "Hybrid trajectory planning for autonomous vehicles using neural networks," in *18th IEEE International Symposium on Computational Intelligence and Informatics (CINTI)*, Budapest, Hungary, 2018, pp. 25-30.

[39] N. Kehtarnavaz, N. Griswold, K. Miller, and P. Lescoe, "A transportable neural-network approach to autonomous vehicle following," *IEEE T Veh. Technol.*, vol. 47, no. 2, pp. 694-702, May 1998.

[40] M. Raj, V. G. Narendra, and Ieee, "Deep neural network approach for navigation of autonomous vehicles," in *6th International Conference for Convergence in Technology (I2CT)*, Electr Network, 2021.

[41] L. Rivals, "Modeling and control of an autonomous vehicle using neural networks," *Mec. Ind. Mater.*, vol. 52, no. 1, pp. 10-12, Mar 1999.

[42] J. Nie, J. Yan, H. L. Yin, L. Ren, and Q. Meng, "A multimodality fusion deep neural network and safety test strategy for intelligent vehicles," *IEEE T Intell. Veh.*, vol. 6, no. 2, pp. 310-322, Jun 2021. DOI: 10.1109/tiv.2020.3027319.

[43] A. Ibrahim, D. Goswami, H. Li, I. M. Soroa, and T. Basten, "Multi-layer multi-rate model predictive control for vehicle platooning under IEEE 802.11p," *Transp. Res. C: Emerg. Technol.*, vol. 124, Mar 2021, Art no. 102905.

[44] H. Zhou, R. Seigel, F. Dion, and L. Yang, "Vehicle platoon control in high-latency wireless communications environment model predictive control method," *Transp. Res. Rec.*, no. 2324, pp. 81-90, 2012.

[45] M. Hu, J. Li, Y. Bian, J. Wang, B. Xu, and Y. Zhu, "Distributed coordinated brake control for longitudinal collision avoidance of multiple connected automated vehicles," *IEEE T Intell. Veh.*, Aug 2022. DOI: 10.1109/TIV.2022.3197951.

[46] J. E. Alves Dias, G. A. Silva Pereira, and R. M. Palhares, "Longitudinal model identification and velocity control of an autonomous car," *IEEE T Intell. Transp. Sys.*, vol. 16, no. 2, pp. 776-786, Apr 2015.

[47] R. E. Stern *et al.*, "Dissipation of stop-and-go waves via control of autonomous vehicles: Field experiments," *Transp. Res. C: Emerg. Technol.*, vol. 89, pp. 205-221, Apr 2018.

[48] H. Zhou, A. Zhou, T. Li, D. Chen, S. Peeta, and J. Laval, "Impact of the low-level controller on string stability of adaptive cruise control system," *arXiv preprint,* arXiv: 2104.07726, 2021.

[49] C. Y. Liang and H. Peng, "Optimal adaptive cruise control with guaranteed string stability," *Veh. Syst. Dyn.*, vol. 32, no. 4-5, pp. 313-330, Nov 1999.

[50] M. Wang, "Infrastructure assisted adaptive driving to stabilise heterogeneous vehicle strings," *Transp. Res. C: Emerg. Technol.*, vol. 91, pp. 276-295, Jun 2018.

[51] S. Wen, G. Guo, B. Chen, and X. Gao, "Cooperative adaptive cruise control of vehicles using a resource-efficient communication mechanism," *IEEE T Intell. Veh.*, vol. 4, no. 1, pp. 127-140, 2019.

[52] J. Sun, Z. Zheng, and J. Sun, "Stability analysis methods and their applicability to car-following models in conventional and connected environments," *Transp. Res. B: Method.*, vol. 109, pp. 212-237, Mar 2018.

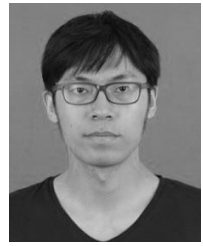

**Hua-Qing Liu** received the B.E. degree in Ship and Ocean Engineering from Harbin Engineering University, Harbin, China, in 2013. He received the Master's degree in Ship and Ocean Engineering from Ningbo University, Ningbo, China, in 2016. He is currently pursuing the Ph.D. degree with the Key Laboratory of Transport Industry of Big Data Application Technologies for Comprehensive Transport, Ministry of Transport, Beijing Jiaotong University. His current research interests include traffic stability analysis and merging algorithm.

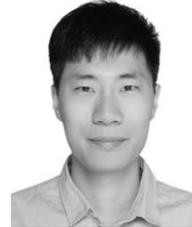

**Shi-Teng Zheng** received the B.E. degree in Traffic and Transportation from Shijiazhuang Tiedao University, Shijiazhuang, China, in 2017. He is currently pursuing the Ph.D. degree with the Key Laboratory of Transport Industry of Big Data Application Technologies for Comprehensive Transport, Ministry of Transport, Beijing Jiaotong University, Beijing, China. His current research interests include traffic flow theory and automated vehicle trajectory planning.

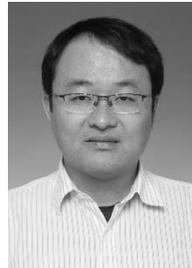

**Rui Jiang** received the B.E. and Ph.D. degrees from the University of Science and Technology of China, Hefei, China, in 1998 and 2003, respectively. He was an Alexander von Humboldt Research Fellow from 2005 to 2006 and a Japanese Society for Promotion of Science Research Fellow from 2008 to 2009. He is currently a Professor with the School of Traffic and Transportation, Beijing Jiaotong University, Beijing, China, where he is involved in the fields of traffic flow theory and intelligent transportation systems.

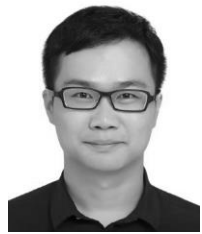

**Junfang Tian** received the Ph.D. degree from Beijing Jiaotong University, Beijing, China, in 2014. He is currently an Associate Professor with the Institute of Systems Engineering, Department of Management and Economics, Tianjin University, China. His current research interests include traffic system simulation and behavior experiments.

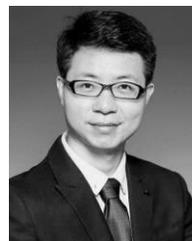

**Ruidong Yan** received the Ph.D. degree from the School of Instrumentation and Optoelectronic Engineering, Beihang University (BUAA), China, in 2017. He was a Post-Doctoral Researcher with Tsinghua University, China. He is currently a Lecturer with the School of Traffic and Transportation, Beijing Jiaotong University, China. His research interests include intelligent transportation systems, intelligent vehicle, reinforcement learning, and anti-disturbance control.

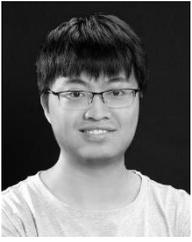

**Fang Zhang** received the B.S. degree in automotive engineering from Tsinghua University, in 2012, and the Ph.D. degree in mechanical engineering from Tsinghua University, in 2018. He joined Beijing Idriverplus Technology Co., Ltd. in 2017. He is currently the technical leader with Beijing Idriverplus Technology Co., Ltd. His research interests include autonomous driving, motion plan, and artificial intelligence.

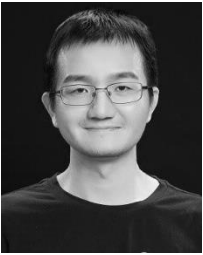

**Dezhao Zhang** received the B.S. degree in automotive engineering from Tsinghua University, in 2006, and the Ph.D. degree in mechanical engineering from Tsinghua University, in 2011. He joined Beijing Idriverplus Technology Co., Ltd. in 2015. He is currently the CEO with Beijing Idriverplus Technology Co., Ltd. His research interests include autonomous driving, computer vision, and vehicle dynamics control.